\documentclass[aps,prd,superscriptaddress,twoside,twocolumn,nofootinbib,showpacs,floatfix]{revtex4-1}
\usepackage{amsmath,amssymb}
\usepackage{graphicx}
\usepackage{subfigure}
\pdfpagewidth=\paperwidth
\pdfpageheight=\paperheight
\allowdisplaybreaks
\newcommand*{\slashed}[1]{{#1\!\!\!/}}
\newcommand*{\hc}{\text{H.\,c.}}
\newcommand{\RNum}[1]{\uppercase\expandafter{\romannumeral #1\relax}}
\usepackage{color}

\begin{document}

\title{\boldmath Photoproduction $\gamma p \to K^+\Lambda(1690)$ in an effective Lagrangian approach}

\author{Neng-Chang Wei}
\affiliation{School of Nuclear Science and Technology, University of Chinese Academy of Sciences, Beijing 101408, China}

\author{Ai-Chao Wang}
\affiliation{School of Nuclear Science and Technology, University of Chinese Academy of Sciences, Beijing 101408, China}

\author{Fei Huang}
\email[Corresponding author. Email: ]{huangfei@ucas.ac.cn}
\affiliation{School of Nuclear Science and Technology, University of Chinese Academy of Sciences, Beijing 101408, China}

\date{\today}

\begin{abstract}
A gauge-invariant model is constructed for the $\gamma p \to K^+\Lambda(1690)$ reaction within a tree-level effective Lagrangian approach with the purpose to understand the underlying production mechanisms and to study the resonance contributions in this reaction. In addition to the $t$-channel $K$ and $K^\ast$ exchanges, the $s$-channel nucleon exchange, and the interaction current, the $s$-channel nucleon resonance exchanges are also included in constructing the reaction amplitudes to describe the data. It is found that the contributions from the $s$-channel $N(2570)5/2^-$ exchange are required to describe the most recently measured total cross-section data for $\gamma p \to K^+\Lambda(1690)$ from the CLAS Collaboration. Further analysis shows that the interaction current dominates the $\gamma p \to K^+\Lambda(1690)$ reaction near the threshold as a result of gauge invariance. The $t$-channel $K$ exchange contributes significantly, while the contributions from the $t$-channel $K^\ast$ exchange as well as the $s$-channel nucleon exchange turn out to be negligible. The contributions from the $s$-channel $N(2570)5/2^-$ exchange are found to be responsible for the bump structure shown in the CLAS total cross-section data above the center-of-mass energy $W \approx 2.7$ GeV. The predictions of the differential cross sections for $\gamma p \to K^+\Lambda(1690)$ are shown and discussed, which can provide theoretical guidances for the future experiments.
\end{abstract}



\maketitle

\section{Introduction}   \label{Sec:intro}

The study of the internal structure of nucleon as well as the spectrum and structures of nucleon resonances gives access to the strong interaction and provides insight into the non-perturbative nature of the quantum chromodynamics (QCD). Today, most of our knowledge of the experimentally established nucleon resonances listed in the Review of Particle Physics (RPP) \cite{Zyla:2020zbs} is mainly coming from the $\pi N$ scattering or $\pi N$, $\eta N$, $K\Lambda$, and $K\Sigma$ photoproduction experiments. However, there might be nucleon resonances that have rather small couplings to the $\pi N$, $\eta N$, $K\Lambda$, and $K\Sigma$ channels, and thus are ``missing'' in these experiments. Actually, both the quark model \cite{Isgur:1977ef,Koniuk:1979vy} and lattice QCD \cite{Edwards:2011jj,Edwards:2012fx} calculations predict significantly more nucleon resonances than those found in experiments. Inspired by this situation, in recent years,  the modern electromagnetic facilities around the world, such as the Thomas Jefferson National
Accelerator Facility (JLab), the Mainz Microtron (MAMI), the 8 GeV Super Photon Ring (SPring-8), and the Electron Stretcher Accelerator (ELSA), have measured large amounts of the data on both differential cross sections and polarization observables for $\eta' N$, $\omega N$, $\phi N$, $K^\ast \Lambda$, $K^\ast \Sigma$, $K\Lambda^\ast$, and $K\Sigma^\ast$ photoproduction processes. These measurements complement the results obtained from the $\pi N$ scattering and $\pi N$, $\eta N$, $K\Lambda$, and $K\Sigma$ photoproduction experiments by providing an alternative platform for finding new nucleon resonances as well as identifying the properties of some known nucleon resonances.

The photoproduction of $K\Lambda^\ast$, with $\Lambda^\ast = \Lambda(1405)$, $\Lambda(1520)$, $\Lambda(1670)$, or $\Lambda(1690)$, is suitable to study nucleon resonances with sizable hidden $s\bar{s}$ components in a relatively less-explored higher energy region due to the much higher reaction thresholds of these channels. Besides, the $s$-channel isospin $I=3/2$ $\Delta$ and $\Delta^\ast$ exchanges are forbidden to contribute in the $K \Lambda^\ast$ photoproduction reactions, which simplifies the reaction mechanisms and facilitates the extraction of the information on the isospin $I=1/2$ nucleon resonances. Compared with the photoproduction of $K\Lambda$ which has been widely studied in various approaches, e.g. chiral perturbation theory \cite{Steininger1997}, isobar models \cite{Luthfiyah:2021yqe,Clymton:2021wof,Kim:2018qfu}, $K$-matrix approaches \cite{Anisovich2017,Cao2013,Hunt2019}, and dynamical coupled-channels models \cite{Kamano2016,Ronchen2018}, the photoproduction of $K \Lambda^\ast$ has also the advantage that it may couple weekly to $\pi N$ and $\eta N$ channels due to its much higher threshold, leading to a much clearer background in isolating the complicated resonance contributions. In literature, intensive experimental and theoretical analyses have been devoted to the investigations of the $K \Lambda^\ast$ photoproduction reactions. 

In Refs.~\cite{Wang:2016dtb,Kim:2017nxg,Zhang:2021iez}, the $K \Lambda(1405)$ photoproduction has been studied   based on the recent differential cross-section data from the CLAS Collaboration \cite{Moriya:2013hwg}. In Refs.~\cite{Nam:2005uq,Nam:2006cx,Nam:2009cv,Nam:2010au,Toki:2007ab,Xie:2010yk,Xie:2013mua,Wang:2014jxb,He:2012ud,He:2014gga,Yu:2017kng,Wei:2021qnc}, the $K \Lambda(1520)$ photoproduction has been studied based on the recent measurements from the CLAS \cite{Moriya:2013hwg}, LEPS \cite{Kohri:2009xe} and SAPHIR \cite{Wieland:2010cq} Collaborations as well as the previous measurements at SLAC \cite{Boyarski:1970yc} and measurements from the LAPS2 group \cite{Barber:1980zv}. Unlike the photoproductions of $K \Lambda(1405)$ and $K \Lambda(1520)$ which have been adequately investigated, theoretical investigations of the $K\Lambda(1670)$ and $K\Lambda(1690)$ photoproductions have been absent in the past due to the lack of experimental data. Fortunately, the total cross-section data for $K\Lambda(1670)$ and $K\Lambda(1690)$ photoproduction reactions from the CLAS Collaboration became available most recently \cite{Shrestha:2021yia}. Theoretical analysis of these data are called on to understand the underlying production mechanisms of these reactions.

In this work, we concentrate on the investigation of the $\gamma p \to K^+\Lambda(1690)$ reaction. We construct a gauge-invariant model within the tree-level effective Lagrangian approach. The goal is to provide a theoretical description of the most recent total cross-section data from the CLAS Collaboration \cite{Shrestha:2021yia}. By doing so, we try to understand the reaction mechanisms of this reaction and, in particular, extract nucleon resonance contributions required in this reaction. We construct the reaction amplitudes for $\gamma p \to K^+\Lambda(1690)$ by including the $t$-channel $K$ and $K^\ast$ exchanges, the $s$-channel nucleon and nucleon resonance exchanges, and the interaction current, with the last one being built in such a way that the full photoproduction amplitude satisfies the generalized Ward-Takahashi identity (WTI) and thus is fully gauge invariant. The results show that the available total cross-section data for $\gamma p \to K^+\Lambda(1690)$ can be well reproduced by considering the $s$-channel $N(2570)5/2^-$ exchange. The contributions from individual Feynman diagrams are analyzed. The differential cross sections are predicted and discussed, which can provide theoretical guidances for the future experiments.

We organize the present paper as follows. In Sec.~\ref{Sec:formalism}, the framework of our theoretical model is briefly introduced. The effective Lagrangians, resonance propagators, and the phenomenological form factors adopted in the present calculations are given explicitly in this section. The numerical results are shown and discussed in Sec.~\ref{Sec:results}. Finally, the summary and conclusions of the present paper are given in Sec.~\ref{sec:summary}.

\section{Formalism}  \label{Sec:formalism}

In this paper, we study the $\gamma p \to K^+\Lambda(1690)$ reaction in an effective Lagrangian approach at the tree-level approximation. For the convenience of further discussion, we define the Mandelstam variables $t=(p-p^\prime)^2=(k-q)^2$, $s=(p+k)^2=(q+p^\prime)^2=W^2$ and $u=(p-q)^2=(p^\prime-k)^2$, with $k$, $q$, $p$ and $p^\prime$ denoting the four-momenta of the incoming photon, outgoing $K$, initial-state proton, and final-state $\Lambda(1690)$, respectively.

\begin{figure}[tbp]
\centering
{\vglue 0.15cm}
\subfigure[~$s$ channel]{
\includegraphics[width=0.45\columnwidth]{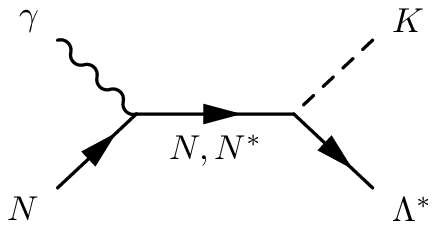}}  {\hglue 0.4cm}
\subfigure[~$t$ channel]{
\includegraphics[width=0.45\columnwidth]{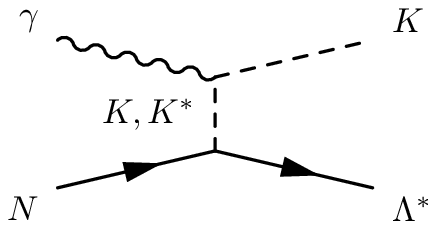}} \\[6pt]
\subfigure[~Interaction current]{
\includegraphics[width=0.45\columnwidth]{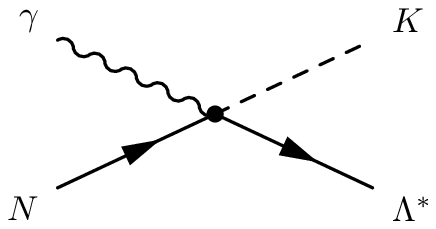}}
\caption{Generic structure of the amplitude for $\gamma p \to K^+\Lambda(1690)$. Time proceeds from left to right. The outgoing $\Lambda^\ast$ denotes $\Lambda(1690)$.}
\label{FIG:feymans}
\end{figure}

As depicted in Fig.~\ref{FIG:feymans}, the following diagrams are included in constructing the reaction amplitudes: the $s$-channel nucleon and nucleon resonance exchanges, the $t$-channel $K$ and $K^\ast$ exchanges, and the interaction current. Note that the $u$-channel $\Lambda$ exchange is not considered as no information on the radial decay $\Lambda(1690) \to \Lambda \gamma$ is available in RPP \cite{Zyla:2020zbs}. The full amputated reaction amplitude for the $\gamma p \to K^+\Lambda(1690)$ reaction can be expressed as 
\begin{equation}
M^{\nu\mu} \equiv M^{\nu\mu}_s + M^{\nu\mu}_t + M^{\nu\mu}_{\rm int},  \label{eq:amplitude}
\end{equation}
with $\mu$ and $\nu$ denoting the Lorentz indices of the incoming photon and outgoing $\Lambda(1690)$, respectively. $M^{\nu\mu}_s$ represents the $s$-channel amplitude of the nucleon and nucleon resonance exchanges, and $M^{\nu\mu}_t$ is the $t$-channel amplitude consisting of the $K$ and $K^\ast$ exchanges. They can be calculated straightforwardly with the effective Lagrangians, resonance propagators, and form factors given in the following part of this section. $M^{\nu\mu}_{\rm int}$ in Eq.~(\ref{eq:amplitude}) represents the generalized interaction current arising from the photon attaching to the internal structure of the $\Lambda(1690)NK$ vertex. It is known that in the effective Lagrangian approach, introducing the phenomenological form factors will result in violation of the gauge invariance of the reaction amplitudes. In the present work, we follow Refs.~\cite{Haberzettl:1997,Haberzettl:2006,Haberzettl:2011zr,Huang:2012,Huang:2013,Wang:2017tpe,Wang:2018vlv,Wei:2019imo,Wang:2020mdn,Wei:2020fmh} to introduce an auxiliary current $C^\mu$ to compensate the violation of the gauge invariance caused by the form factors and model the generalized interaction current $M^{\nu\mu}_{\rm int}$ as
\begin{equation}
M^{\nu\mu}_{\rm int} = \Gamma_{\Lambda^\ast NK}^\nu(q) C^\mu + M^{\nu\mu}_{\rm KR} f_t,   \label{eq:Mint}
\end{equation}
where $\Gamma_{\Lambda^\ast NK}^\nu(q)$ is the vertex function of the $\Lambda(1690)NK$ interaction obtained from the Lagrangian of Eq.~(\ref{eq:lpls}),
\begin{equation}
\Gamma_{\Lambda^\ast NK}^\nu(q) = - \frac{g_{\Lambda^\ast NK}}{M_K} \gamma_5 q^\nu,
\end{equation}
and $M^{\nu\mu}_{\rm KR}$ is the traditional Kroll-Ruderman term obtained from the Lagrangian of Eq.~(\ref{eq:con}),
\begin{equation}
M^{\nu\mu}_{\rm KR} = \frac{g_{\Lambda^\ast NK}}{M_K} g^{\nu\mu} \gamma_5 Q_K \tau,
\end{equation}
with $Q_K$ being the electric charge of the outgoing $K$ meson and $\tau$ being the isospin factor of the Kroll-Ruderman term. $f_t$ is the phenomenological form factor attaching to the amplitude of the $t$-channel $K$ exchange, which is given by Eq.~(\ref{eq:ff_M}). The auxiliary current $C^\mu$ for the $\gamma p \to K^+\Lambda(1690)$ reaction is chosen to be \cite{Haberzettl:2006,Haberzettl:2011zr,Huang:2012}
\begin{equation}
C^\mu =  - Q_K \tau \frac{f_t-\hat{F}}{t-q^2}  (2q-k)^\mu - \tau Q_N \frac{f_s-\hat{F}}{s-p^2} (2p+k)^\mu,  \label{eq:GI-auxi}
\end{equation}
with
\begin{equation} \label{eq:Fhat}
\hat{F} = 1 - \hat{h} \left(1 -  f_s\right) \left(1 - f_t\right).
\end{equation}
Here $Q_N$ represents the electric charge of $N$ and $f_s$ is the phenomenological form factor for the $s$-channel $N$ exchange as given in Eq.~(\ref{eq:ff_B}); $\hat{h}$ is an arbitrary function going to unity in high-energy limit and set to be $1$ in the present work for simplicity. 

The prescriptions of the generalized interaction current in Eq.~(\ref{eq:Mint}) and the auxiliary current $C^\mu$ in Eq.~(\ref{eq:GI-auxi}) ensure that the full photoproduction amplitude of Eq.~(\ref{eq:amplitude}) satisfies the generalized WTI and thus is fully gauge invariant \cite{Haberzettl:2006,Haberzettl:2011zr,Huang:2012}, independent of what particular forms are chosen for the phenomenological form factors $f_s$ and $f_t$.

\subsection{Effective Lagrangians} \label{Sec:Lagrangians}

In this subsection, the Lagrangians needed to calculate the reaction amplitudes of the $\gamma p \to K^+\Lambda(1690)$ reaction are given. For convenience, we define the following operators:
\begin{equation}
\Gamma^{(+)}=\gamma_5 \qquad {\rm and} \qquad \Gamma^{(-)}=1,
\end{equation}
and the field-strength tensor for the photon field $A^\mu$:
\begin{equation}
F^{\mu\nu} = \partial^\mu A^\nu - \partial^\nu A^\mu.
\end{equation}
To avoid any ambiguities, the notation $\Lambda^\ast$ will stand for only the $\Lambda(1690)$ resonance in the rest of the present paper.

We use the following Lagrangians to calculate the non-resonant amplitudes
\begin{eqnarray}
{\cal L}_{\gamma NN} &=& -\,e \bar{N} \!\left[ \! \left( \hat{e} \gamma^\mu - \frac{ \hat{\kappa}_N} {2M_N}\sigma^{\mu \nu}\partial_\nu \! \right) \! A_\mu\right]\! N, \\[6pt]
{\cal L}_{\gamma KK} &=& ie \!\left[K^+\left(\partial_\mu K^-\right)-K^-\left(\partial_\mu K^+\right)\right] \! A^\mu, \\[6pt]
{\cal L}_{\gamma K{K^\ast}} &=& e\frac{g_{\gamma K{K^\ast}}}{M_K}\varepsilon^{\alpha \mu \lambda \nu}\left(\partial_\alpha A_\mu\right)\left(\partial_\lambda K\right) K^\ast_\nu, \label{Lag:gKKst}   \\[6pt]
{\cal L}_{\gamma \Lambda^\ast NK} &=& -iQ_K\frac{g_{\Lambda^\ast NK}}{M_K} \bar{\Lambda}^{\ast \mu} A_\mu K\gamma_5 N + \hc, \label{eq:con}  \\[6pt]
{\cal L}_{\Lambda^\ast NK} &=& \frac{g_{\Lambda^\ast NK}}{M_K} {\bar{\Lambda}}^{\ast \mu} \left(\partial_\mu K\right)\gamma_5 N + \hc, \label{eq:lpls} \\[6pt]
{\cal L}_{\Lambda^\ast N K^\ast} &=& -\frac{i g_{\Lambda^\ast N K^\ast}}{M_{K^\ast}}\bar{\Lambda}^{\ast \mu}\gamma^\nu \left(\partial_\mu K^{\ast}_\nu - \partial_\nu K^{\ast}_\mu \right)N  \nonumber \\
&& + \, \hc, 
\end{eqnarray}
with $M_N$, $M_K$, and $M_{K^\ast}$ denoting the masses of $N$, $K$, and $K^\ast$, respectively. $e$ is the elementary charge unit and $\hat{e}$ stands for the charge operator acting on the nucleon field. $\hat{\kappa}_N \equiv \kappa_p\hat{e} + \kappa_n(1-\hat{e})$ with $\kappa_p=1.793$ being the anomalous magnetic moment of proton and $\kappa_n=-1.913$ the anomalous magnetic moment of neutron. The coupling constants $g_{\gamma K K^{\ast}} = 0.413$ and $g_{\Lambda^\ast NK} =4.20$ are calculated from the decay width of $\Gamma(K^{*\pm}\to K^{\pm}\gamma) \approx 0.0503$ MeV and $\Gamma(\Lambda(1690)\to N\bar{K}) \approx 17.5$ MeV, respectively, as given by RPP \cite{Zyla:2020zbs}. The coupling $g_{\Lambda^\ast N K^\ast}$ is treated as a free parameter to be determined by a fit to the data.

For the $s$-channel resonance exchanges, the Lagrangians for the electromagnetic interactions read \cite{Wei:2021qnc,Wang:2017tpe,Wang:2018vlv,Wei:2019imo,Wang:2020mdn}
\begin{eqnarray}
{\cal L}_{RN\gamma}^{1/2\pm} &=& e\frac{g_{RN\gamma}^{(1)}}{2M_N}\bar{R} \Gamma^{(\mp)}\sigma_{\mu\nu} \left(\partial^\nu A^\mu \right) N  + \hc, \\[6pt]
{\cal L}_{RN\gamma}^{3/2\pm} &=& -\, ie\frac{g_{RN\gamma}^{(1)}}{2M_N}\bar{R}_\mu \gamma_\nu \Gamma^{(\pm)}F^{\mu\nu}N \nonumber \\
&&+\, e\frac{g_{RN\gamma}^{(2)}}{\left(2M_N\right)^2}\bar{R}_\mu \Gamma^{(\pm)}F^{\mu \nu}\partial_\nu N + \hc, \\[6pt]
{\cal L}_{RN\gamma}^{5/2\pm} & = & e\frac{g_{RN\gamma}^{(1)}}{\left(2M_N\right)^2}\bar{R}_{\mu \alpha}\gamma_\nu \Gamma^{(\mp)}\left(\partial^{\alpha} F^{\mu \nu}\right)N \nonumber \\
&& \pm\, ie\frac{g_{RN\gamma}^{(2)}}{\left(2M_N\right)^3}\bar{R}_{\mu \alpha} \Gamma^{(\mp)}\left(\partial^\alpha F^{\mu \nu}\right)\partial_\nu N \nonumber \\
&& + \,  \hc,  \\[6pt]
{\cal L}_{RN\gamma}^{7/2\pm} &=&  ie\frac{g_{RN\gamma}^{(1)}}{\left(2M_N\right)^3}\bar{R}_{\mu \alpha \beta}\gamma_\nu \Gamma^{(\pm)}\left(\partial^{\alpha}\partial^{\beta} F^{\mu \nu}\right)N \nonumber \\
&&-\, e\frac{g_{RN\gamma}^{(2)}}{\left(2M_N\right)^4}\bar{R}_{\mu \alpha \beta} \Gamma^{(\pm)} \left(\partial^\alpha \partial^\beta F^{\mu \nu}\right) \partial_\nu N  \nonumber \\
&&  + \,  \hc,
\end{eqnarray}
and the Lagrangians for resonances coupling to $\Lambda(1690)K$ are \cite{Wei:2021qnc}
\begin{eqnarray}
{\cal L}_{R\Lambda^\ast K}^{1/2\pm} &=& \frac{g^{(1)}_{R\Lambda^\ast K}}{M_K} \bar{\Lambda}^{\ast\mu}\Gamma^{(\pm)} \left(\partial_\mu K\right) R + \hc,  \\[6pt]
{\cal L}_{R\Lambda^\ast K}^{3/2\pm} &=& \frac{g^{(1)}_{R\Lambda^\ast K}}{M_K} \bar{\Lambda}^{\ast\mu}\gamma_\nu \Gamma^{(\mp)} \left(\partial^\nu K\right) R_\mu \nonumber \\
&& + \, i \frac{g^{(2)}_{R\Lambda^\ast K}}{M_K^2} \bar{\Lambda}^\ast_\alpha \Gamma^{(\mp)} \left(\partial^\mu \partial^\alpha K\right) R_\mu  + \hc, \\[6pt]
{\cal L}_{R\Lambda^\ast K}^{5/2\pm} &=& i\frac{g^{(1)}_{R\Lambda^\ast K}}{M_K^2}  \bar{\Lambda}^{\ast\alpha}\gamma_\mu \Gamma^{(\pm)} \left(\partial^\mu \partial^\beta K\right) R_{\alpha\beta} \nonumber  \\
&&- \,\frac{g^{(2)}_{R\Lambda^\ast K}}{M_K^3} \bar{\Lambda}^\ast_\mu \Gamma^{(\pm)} \left(\partial^\mu \partial^\alpha \partial^\beta K\right) R_{\alpha\beta} \nonumber \\
&& + \, \hc, \\[6pt]
{\cal L}_{R\Lambda^\ast K}^{7/2\pm} &=& -\frac{g^{(1)}_{R\Lambda^\ast K}}{M_K^3} \bar{\Lambda}^{\ast\alpha}\gamma_\mu \Gamma^{(\mp)} \left(\partial^\mu \partial^\beta \partial^\lambda K\right) R_{\alpha\beta\lambda} \nonumber \\
&& - \, i \frac{g^{(2)}_{R\Lambda^\ast K}}{M_K^4} \bar{\Lambda}^{\ast}_\mu \Gamma^{(\mp)} \left(\partial^\mu \partial^\alpha \partial^\beta \partial^\lambda K\right) R_{\alpha\beta\lambda} \nonumber \\
&& + \, \hc,
\end{eqnarray}
with $R$ designating the $N^\ast$ resonance and the superscript of ${\cal L}_{RN\gamma}$ and ${\cal L}_{R\Lambda^\ast K}$ being the spin and parity of the resonance $R$. In the present work, the $g^{(2)}_{R\Lambda^\ast K}$ terms in ${\cal L}_{R\Lambda^\ast K}^{3/2\pm}$, ${\cal L}_{R\Lambda^\ast K}^{5/2\pm}$, and ${\cal L}_{R\Lambda^\ast K}^{7/2\pm}$ are ignored for the sake of simplicity. The products of the coupling constants $g_{RN\gamma}^{(i)} g^{(1)}_{R\Lambda^\ast K}$ ($i=1,2$) are treated as fit parameters.

\subsection{Resonance propagators}

The propagators of the $N^\ast$ resonance field ($R$) with mass $M_R$, width $\Gamma_R$, four-momentum $p$, and spin $1/2$, $3/2$, $5/2$, and $7/2$ are expressed as \cite{Wang:2017tpe,Behrends:1957,Fronsdal:1958,Zhu:1999} \begin{eqnarray}
S_{1/2}(p) &=& \frac{i}{\slashed{p} - M_R + i \Gamma_R/2}, \label{propagator-1hf}  \\[6pt]
S_{3/2}(p) &=&  \frac{i}{\slashed{p} - M_R + i \Gamma_R/2} \left( \tilde{g}_{\mu \nu} + \frac{1}{3} \tilde{\gamma}_\mu \tilde{\gamma}_\nu \right),  \label{propagator-3hf} \\[6pt]
S_{5/2}(p) &=&  \frac{i}{\slashed{p} - M_R + i \Gamma_R/2} \,\bigg[ \, \frac{1}{2} \big(\tilde{g}_{\mu \alpha} \tilde{g}_{\nu \beta} + \tilde{g}_{\mu \beta} \tilde{g}_{\nu \alpha} \big)  \nonumber \\
&& -\, \frac{1}{5}\tilde{g}_{\mu \nu}\tilde{g}_{\alpha \beta}  + \frac{1}{10} \big(\tilde{g}_{\mu \alpha}\tilde{\gamma}_{\nu} \tilde{\gamma}_{\beta} + \tilde{g}_{\mu \beta}\tilde{\gamma}_{\nu} \tilde{\gamma}_{\alpha}  \nonumber \\
&& +\, \tilde{g}_{\nu \alpha}\tilde{\gamma}_{\mu} \tilde{\gamma}_{\beta} +\tilde{g}_{\nu \beta}\tilde{\gamma}_{\mu} \tilde{\gamma}_{\alpha} \big) \bigg], \\[6pt]
S_{7/2}(p) &=&  \frac{i}{\slashed{p} - M_R + i \Gamma_R/2} \, \frac{1}{36}\sum_{P_{\mu} P_{\nu}} \bigg( \tilde{g}_{\mu_1 \nu_1}\tilde{g}_{\mu_2 \nu_2}\tilde{g}_{\mu_3 \nu_3} \nonumber \\
&& -\, \frac{3}{7}\tilde{g}_{\mu_1 \mu_2}\tilde{g}_{\nu_1 \nu_2}\tilde{g}_{\mu_3 \nu_3} + \frac{3}{7}\tilde{\gamma}_{\mu_1} \tilde{\gamma}_{\nu_1} \tilde{g}_{\mu_2 \nu_2}\tilde{g}_{\mu_3 \nu_3} \nonumber \\
&& -\, \frac{3}{35}\tilde{\gamma}_{\mu_1} \tilde{\gamma}_{\nu_1} \tilde{g}_{\mu_2 \mu_3}\tilde{g}_{\nu_2 \nu_3} \bigg),  \label{propagator-7hf}
\end{eqnarray}
where
\begin{eqnarray}
\tilde{g}_{\mu \nu} &=& -\, g_{\mu \nu} + \frac{p_{\mu} p_{\nu}}{M_R^2}, \\[6pt]
\tilde{\gamma}_{\mu} &=& \gamma^{\nu} \tilde{g}_{\nu \mu} = -\gamma_{\mu} + \frac{p_{\mu}\slashed{p}}{M_R^2},   \label{eq:prop-auxi}
\end{eqnarray}
and the summation over $P_\mu \left(P_\nu\right)$ in Eq.~(\ref{propagator-7hf}) goes over the $3!=6$ possible permutations of the indices $\mu_1\mu_2\mu_3$ $\left(\nu_1\nu_2\nu_3\right)$.

In the present work the Rarita-Schwinger prescriptions for propagators of high spin resonances are used in Eqs.~(\ref{propagator-3hf})-(\ref{propagator-7hf}). It is known that the Rarita-Schwinger prescriptions suffer from the low-spin background problems. The efforts to find the pure high-spin propagator formalism was started in Ref.~\cite{DelgadoAcosta:2015ypa}, and later was implemented for use in nuclear and particle physics in Ref.~\cite{Kristiano:2017qjq}. In future when more data for $\gamma p \to K^+ \Lambda(1690)$ become available, a serious treatment of the resonance propagators as discussed in Refs.~\cite{DelgadoAcosta:2015ypa, Kristiano:2017qjq} will be needed.

\subsection{Form factors} \label{subSec:form factor}

In the effective Lagrangian approach, phenomenological form factors are introduced in hadronic vertices to taken into account the internal structures of hadrons and to regularize the momentum dependence of the reaction amplitudes. In the present work, we use the following form factors for the $t$-channel meson exchanges \cite{Wei:2021qnc,Wang:2017tpe,Wang:2020mdn,Wang:2018vlv,Wei:2019imo}
\begin{eqnarray}
f_M(q^2_M) =  \left (\frac{\Lambda_M^2-M_M^2}{\Lambda_M^2-q^2_M} \right)^2,   \label{eq:ff_M}
\end{eqnarray}
and for the $s$-channel baryon exchanges, we use \cite{Wei:2021qnc,Wang:2017tpe,Wang:2020mdn,Wang:2018vlv,Wei:2019imo}
\begin{eqnarray}
f_B(p^2_s) =\left (\frac{\Lambda_B^4}{\Lambda_B^4+\left(p_s^2-M_B^2\right)^2} \right )^2,  \label{eq:ff_B}
\end{eqnarray}
with $q_M$ standing for the four-momentum of the intermediate meson in the $t$ channel, $p_s$ standing for the four-momentum of the intermediate baryon in the $s$ channel, and $\Lambda_{M(B)}$ being the cutoff parameter. In the present work, we adopt a common value for all the non-resonant diagrams, i.e. $\Lambda_{\rm bg} \equiv \Lambda_K=\Lambda_{K^\ast}=\Lambda_N$, to reduce the number of adjustable parameters. The parameter $\Lambda_{\rm bg}$ and the cutoff parameter $\Lambda_R$ for nucleon resonances are determined by fitting the experimental data.

Besides the dipole form factors as adopted in the present work, the  monopole form factors and exponential (Gaussian) form factors are also widely used in literature. We have checked and found that a monopole type of form factor will result in a much bigger $\chi^2$, indicting a much worse fitting quality, while an exponential (Gaussian) type of form factor will result in similar results as those presented in the present work, although the model parameters change a little bit.

\section{Results and discussion}   \label{Sec:results}

\begin{figure}[tbp]
\includegraphics[width=\columnwidth]{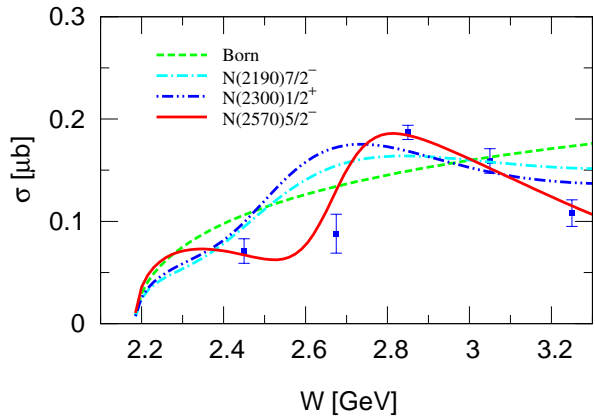}
\caption{Total cross sections for $\gamma p\to K^+\Lambda(1690)$. The green dashed line represents the results without any nucleon resonances. The cyan dot-dashed, blue double-dot-dashed, and red solid lines denote the results obtained by including the $N(2190)7/2^-$, $N(2300)1/2^+$, and  $N(2570)5/2^-$ resonances in the $s$ channel, respectively. Data are taken from the CLAS Collaboration \cite{Shrestha:2021yia}.}  
\label{fig:compare}
\end{figure}

As mentioned in the introduction section of this paper, very recently, the first data on total cross sections for the $\gamma p \to K^+ \Lambda(1690)$ reaction became available from the CLAS Collaboration \cite{Shrestha:2021yia}. In the present work, we for the first time perform a theoretical analysis of these data within an effective Lagrangian approach. We construct the reaction amplitudes by including the $t$-channel $K$ and $K^\ast$ exchanges, the $s$-channel nucleon and nucleon resonance exchanges, and the interaction current. The gauge invariance of the photoproduction amplitudes is fully implemented by introducing an auxiliary current in the generalized contact term (cf. Eq.~(\ref{eq:Mint})). As the data for $\gamma p \to K^+ \Lambda(1690)$ are scarce, the strategy adopted in the present work for introducing nucleon resonances is that we introduce nucleon resonances as few as possible to describe the available data.

We first try to see how the total cross-section data of the $\gamma p\to K^+\Lambda(1690)$ reaction \cite{Shrestha:2021yia} can be described without introducing any nucleon resonances. In Fig.~\ref{fig:compare} we show the results obtained from the Born amplitudes with green dashed line. One sees clearly that, the Born amplitudes themselves are far from sufficient to describe the data, indicating the necessity of the contributions from the resonance exchanges.

In RPP \cite{Zyla:2020zbs}, there are three nucleon resonances lying above the threshold of the $\gamma p\to K^+\Lambda(1690)$ reaction with spin $J \leq 7/2$, namely, the $N(2190)7/2^-$, $N(2300)1/2^+$, and $N(2570)5/2^-$ resonances.\footnote{For resonance with spin $J \geq 9/2$, the vertices and propagators are much more complicated. We postpone the inclusion of them till more data for this reaction become accessible.} We then try to include one of them in the $s$-channel to describe the data. We use Minuit to fit the model parameters, and the resulted $\chi^2$ per data are $7.69$, $7.64$, and $1.17$, respectively, for results including each of the $N(2190)7/2^-$, $N(2300)1/2^+$, and $N(2570)5/2^-$ resonances. The corresponding results are shown in Fig.~\ref{fig:compare}, where the cyan dot-dashed, blue double-dot-dashed, and red solid lines denote the results obtained by including the $N(2190)7/2^-$, $N(2300)1/2^+$, and  $N(2570)5/2^-$ resonances in the $s$ channel, respectively. One can see that, while the fits with the $N(2190)7/2^-$ and  $N(2300)1/2^+$ resonances fail to reproduce the data, the fit with the $N(2570)5/2^-$ resonance describe the data satisfactorily well.

\begin{table}[tb]
\caption{Fitted values of model parameters. The asterisks below the resonance $N(2570){5/2}^-$ denote the overall status of this resonance evaluated by RPP \cite{Zyla:2020zbs}. The values in the brackets below the mass $M_R$ and width $\Gamma_R$ of the resonance $N(2570){5/2}^-$ are the corresponding values listed in RPP \cite{Zyla:2020zbs} which are quoted from the results of BESIII \cite{Ablikim:2012zk}. $\sqrt{\beta_{\Lambda^\ast K}}A_{j}$ is the reduced helicity amplitude for resonance $N(2570)5/2^-$ with $\beta_{\Lambda^\ast K}$ denoting the branching ratio of the resonance decay to $\Lambda(1690) K$ and $A_{j}$ standing for the helicity amplitude with spin $j$ for resonance radiative decay to $\gamma p$.}
\label{table:constants}
\renewcommand{\arraystretch}{1.2}
\begin{tabular*}{\columnwidth}{@{\extracolsep\fill}lc}
\hline\hline
$\Lambda_{\rm bg}$ $[{\rm MeV}]$      &   $965 \pm 16$		\\
$g_{\Lambda^\ast N K^\ast}$      &   $-19.32 \pm 14.76$	\\
\hline
                  			    &   $N(2570){5/2}^-$   	\\
		                        &  $\ast \ast$ \\
$M_R$ $[{\rm MeV}]$             &  $2660 \pm 5$	\\
		                        &  [$2570^{+19+34}_{-10-10}$]			\\
$\Gamma_R$ $[{\rm MeV}]$        &  $300 \pm 16$	 \\
				                &  [$250^{+14+69}_{-24-21}$]	\\
$\Lambda_{R}$ $[{\rm MeV}]$     &  $2390 \pm 82$	\\
$\sqrt{\beta_{\Lambda^\ast K}}A_{1/2}$ $[10^{-3}\,{\rm GeV}^{-1/2}]$  & $-4.61 \pm 0.06$    \\
$\sqrt{\beta_{\Lambda^\ast K}}A_{3/2}$ $[10^{-3}\,{\rm GeV}^{-1/2}]$  & $6.24  \pm 0.23$    \\
\hline\hline
\end{tabular*}
\end{table}

\begin{figure}[tbp]
\includegraphics[width=\columnwidth]{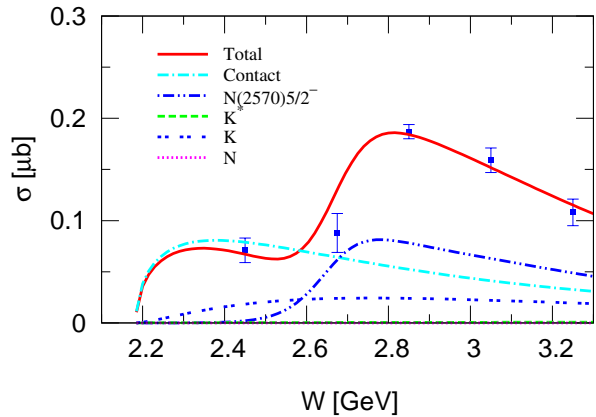}
\caption{Total cross sections for $\gamma p\to K^+\Lambda(1690)$. The red solid line denotes the results obtained by the full reaction amplitude with the $N(2570)5/2^-$ resonance. The cyan dot-dashed, blue double-dot-dashed, green dashed, blue double-dotted, and magenta dotted lines represent the results from the individual contributions of the interaction current, $s$-channel $N(2570)5/2^-$ exchange, $t$-channel $K^\ast$ exchange, $t$-channel $K$ exchange, and $s$-channel $N$ exchange, respectively. Data are taken from the CLAS Collaboration \cite{Shrestha:2021yia}.}  
\label{fig:1690}
\end{figure}

\begin{figure}[tbp]
\center
\includegraphics[width=\columnwidth]{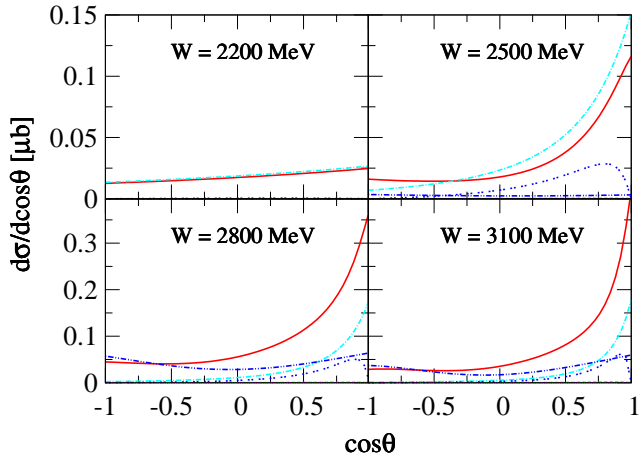}
\caption{Predicted differential cross sections for $\gamma p\to K^+\Lambda(1690)$ at four selected energies. The red solid lines denote the results obtained by the full reaction amplitudes with the $N(2570)5/2^-$ resonance. The cyan dot-dashed, blue double-dot-dashed, green dashed, blue double-dotted, and magenta dotted lines represent the individual contributions from the interaction current, $s$-channel $N(2570)5/2^-$ exchange, $t$-channel $K^\ast$ exchange, $t$-channel $K$ exchange, and $s$-channel $N$ exchange, respectively.}
\label{fig:dsig}
\end{figure}

In rest of this section, we come to a detailed discussion of the model results for the $\gamma p\to K^+\Lambda(1690)$ reaction obtained by including the $N(2570)5/2^-$ resonance. In Table~\ref{table:constants} we list the fitted values of the corresponding model parameters. For resonance couplings, we list the reduced helicity amplitudes $\sqrt{\beta_{\Lambda^\ast K}}A_j$ instead of showing their strong and electromagnetic coupling constants separately \cite{Wei:2021qnc,Wang:2017tpe,Huang:2013,Wang:2019mid,Wei:2020fmh}, since in tree-level calculations as performed in the present work, only the products of the resonance strongnic and electromagnetic coupling constants are relevant to the reaction amplitudes. Here $\beta_{\Lambda^\ast K}$ is the branching ratio for resonance decay to $\Lambda(1690) K$, and $A_j$ is the helicity amplitude with spin $j$ ($j=1/2,3/2$) for resonance radiative decay to $\gamma p$. The values in the brackets below the mass $M_R$ and width $\Gamma_R$ of the resonance $N(2570){5/2}^-$ are the corresponding values listed in RPP \cite{Zyla:2020zbs}, which are quoted from the results of BESIII \cite{Ablikim:2012zk}.  One can see that both the fitted mass and width for the resonance $N(2570)5/2^-$ are close to the ranges given by BESIII \cite{Ablikim:2012zk}.

In Fig.~\ref{fig:1690}, the model results of the total cross sections (the red solid line) for the $\gamma p\to K^+\Lambda(1690)$ reaction are shown and compared to the corresponding data from the CLAS Collaboration \cite{Shrestha:2021yia}. The contributions calculated from each individual reaction amplitudes are also shown with the cyan dot-dashed, blue double-dot-dashed, green dashed, blue double-dotted, and magenta dotted lines representing the contributions from the individual amplitudes of the interaction current, $s$-channel $N(2570)5/2^-$ exchange, $t$-channel $K^\ast$ exchange, $t$-channel $K$ exchange, and $s$-channel $N$ exchange, respectively. One sees that, near the threshold the interaction current dominates the $\gamma p \to K^+ \Lambda(1690)$ reaction as a consequence of gauge invariance. The $t$-channel $K$ exchange makes considerable contributions to the $\gamma p \to K^+ \Lambda(1690)$ reaction and its destructive interference with the interaction current is observed. The contributions from the $t$-channel $K^\ast$ exchange as well as the $s$-channel nucleon exchange are very small. The $N(2570)5/2^-$ resonance exchange provides significant contributions at high energies, and actually, its constructive interference with the interaction current is responsible for the bump structure shown in the CLAS total cross-section data at the center-of-mass energies above $W \approx 2.7$ GeV.

In Fig.~\ref{fig:dsig}, we show the predictions of the differential cross sections for $\gamma p\to K^+\Lambda(1690)$ from the present model. The contributions from individual interaction diagrams are shown in this figure as well. One sees that, at the lower energy region, the interaction current dominates the differential cross sections of this reaction and has destructive interference with the $t$-channel $K$ exchange which contributes mainly at the forward angles. Note that, the dominance of the interaction current is also observed in the $\gamma p \to K^+\Lambda(1520)$ reaction \cite{Xie:2013mua,Wang:2014jxb,He:2014gga,Wei:2021qnc}.  As energy increases, the contributions from the $s$-channel $N(2570)5/2^-$ exchange become prominent. These predicted differential cross sections for the $\gamma p\to K^+\Lambda(1690)$ reaction provide theoretical guidances to the experimental measurements and can be tested by the future data.

\section{Summary and conclusion}  \label{sec:summary}

The first measurement of the total cross sections for the $\gamma p \to K^+ \Lambda(1690)$ reaction was presented most recently by the CLAS Collaboration in Ref.~\cite{Shrestha:2021yia}. In the present work, we for the first time performed a theoretical analysis of these data within an effective Lagrangian approach. Apart from the $s$-channel nucleon exchange, the $t$-channel $K$ and $K^\ast$ exchanges, and the generalized interaction current, we considered as few as possible nucleon resonance exchanges in the $s$ channel in constructing the reaction amplitudes for $\gamma p \to K^+ \Lambda(1690)$ to describe the data. The gauge invariance of the full photoproduction amplitude was fully implemented by considering a particular auxiliary current in the generalized interaction current (cf. Eqs.~(\ref{eq:Mint}) and (\ref{eq:GI-auxi})).

It was found that the available total cross-section data for the $\gamma p \to K^+ \Lambda(1690)$ reaction can be satisfactorily described by introducing the $s$-channel $N(2570)5/2^-$ exchange. The fitted mass and width for $N(2570)5/2^-$ are close to the ranges listed in RPP \cite{Zyla:2020zbs} which are quoted from the results of BESIII \cite{Ablikim:2012zk}. The interaction current was found to dominate the cross sections of $\gamma p \to K^+ \Lambda(1690)$ near the reaction threshold. Considerable contributions from the $t$-channel $K$ exchange were observed. The $t$-channel $K^\ast$ exchange as well as the $s$-channel nucleon exchange were found to make rather small contributions to the cross sections of $\gamma p \to K^+ \Lambda(1690)$. The $N(2570)5/2^-$ resonance exchange provides significant contributions at high energies, and its constructive interference with the interaction current was found to be responsible for the bump structure shown in the CLAS total cross-section data at the center-of-mass energies above $W \approx 2.7$ GeV. The differential cross sections for the $\gamma p \to K^+ \Lambda(1690)$ reaction were predicted, which provide theoretical guidances for the future experiments.

\begin{acknowledgments}
This work is partially supported by the National Natural Science Foundation of China under Grants No.~12175240, No.~11475181, and No.~11635009, the Fundamental Research Funds for the Central Universities, the China Postdoctoral Science Foundation under Grant No.~2021M693141 and No.~2021M693142, and the Key Research Program of Frontier Sciences of Chinese Academy of Sciences under Grant No.~Y7292610K1.
\end{acknowledgments}

\end{document}